\documentclass[twocolumn,preprintnumbers,amsmath,amssymb]{revtex4}
\usepackage{amssymb}
\usepackage{txfonts}
\usepackage{bbm}
\usepackage{graphicx}
\usepackage{appendix}
\usepackage{epsf}

\begin{document}

\title{Quantum phase transition in the Fulde-Ferrell-Larkin-Ovchinnikov states of a
two dimensional d-wave superconductor upon increasing the exchange
field}

\author
{Tao Zhou and C. S. Ting}

\affiliation{Texas Center for Superconductivity and Department of
Physics, University of Houston, Houston, Texas 77204}

\date{\today}

\begin{abstract}
We studied the phase diagram for a two-dimensional $d$-wave
superconducting system under an in-plane magnetic field or an
exchange field. According to the  spatial configuration of the order
parameter, we show that there exists  quantum phase transitions in
which the uniform phase transforms to the one-dimensional
Fulde-Ferrell-Larkin-Ovchinnikov (FFLO) state, and then to
two-dimensional FFLO state upon increasing the exchange field.
 The local density of states are calculated and suggested to be signatures to distinguish these phases.

 \end{abstract}
\maketitle

The Fulde-Ferrell- Larkin-Ovchinnikov (FFLO) state was predicted
several decades ago by Fulde and Ferrell~\cite{ful}, and Larkin and
Ovchinnikov~\cite{lar} for the superconductor in a strong magnetic
field, where the superconducting (SC) order parameter varies
periodically in space. While  the occurance of the FFLO
state requires very stringent conditions on the SC materials,
namely, the Pauli paramagnetism effect should dominate over the
orbital effect~\cite{grue}, and the material needs to be very
clean~\cite{asl}. As a result, this long thought of inhomogeneous SC
state has never been observed in conventional superconductors.

For layered systems with an exchange field or a magnetic field
parallel to the SC plane, the orbital effect will be suppressed
strongly due to the low dimensionality. Thus they could be strong
candidates to look for the FFLO state. Actually, in the past decade,
indications for possible FFLO state have been reported in the heavy
fermion materials CeCoIn$_5$~\cite{bia,kum}, organic superconductors
$\lambda$-(BETS)$_2$GaCl$_4$~\cite{tan},$\lambda$-(BETS)$_2$FeCl$_4$~\cite{uji,bal}
and $\kappa$-(BEDT-TTF)$_2$Cu(NCS)$_2$~\cite{shi,man}. All of them
are quasi two dimensional (2D) layered compounds. The experimental
developments have attracted renewed interest on the property of the
FFLO state. Theoretically, the existence and the character of the
FFLO state can be investigated through analyzing the Ginzburg-Landau
(GL) free-energy function, which is valid at temperatures not too
below  the superconducting transition temperature. Another effective
method is the Bogoliubov-de-Gennes (BdG) technique and it has been
proven to be a powerful tool to study the inhomogeneous state and
the local density of states (LDOS) self-consistently in the
low-dimensional system. In fact, in the past, the FFLO state has
been studied intensively based on the above two
techniques~\cite{maki,qian,cui,qianw,gong,mats,agt}. For a 2D
system, it is somewhat established~\cite{maki,qian}that the order
parameter has a 2D checkerboard pattern for a superconductor with
$d$-wave pairing symmetry, and a 1D stripe-like pattern for $s$-wave
pairing symmetry~\cite{qian,cui}.
 On the other hand, it has been shown that in the presence of dilute impurities,
 the pattern of the FFLO state becomes 1D stripe-like in a 2D $d$-wave superconductor ~\cite{qianw,gong}.
 This implies that the 2D and 1D FFLO states may be present in the system in different parameter region.
 It was also argued without a  calculation that the $H-T$ phase diagram for the 2D isotropic systems, regardless it is $s$ or $d$-wave pairing, should include both 1D FFLO state and 2D FFLO states with
square, triangular and hexagonal patterns~\cite{mats} as the
exchange  interaction becomes stronger.  This result is
significantly different from our previous understanding about the
pattern of the FFLO state in 2D  systems. Therefore it is of
interest and timely to reexamine the formation of the FFLO state in
a 2D superconducting system as a function of the exchange
interaction.

In the present work, we calculate the spatially distributed order
parameter self-consistently based on the BdG equations. The whole
$H-T$ phase diagram is constructed. We verify numerically that the
pattern of order parameters in the FFLO state for two-dimensional
$d$-wave square lattice samples in presence of an exchange field is
not always simply 2D. At zero temperature, the pattern changes from
uniform to 1D FFLO state, and then to 2D FFLO state as the strength
of the exchange field increases. The periodicity of the 1D and 2D
FFLO states decreases as the exchange field increases. At finite
temperature, the 1D FFLO state will transit to the uniform phase
upon increasing the temperature. Thus near the SC transition
temperature, only uniform phase and 2D FFLO state are observed. The
LDOS for the above states are also calculated and they provide
definitive signatures for the above mentioned FFLO states. In
addition, our numerical study indicates that for a 2D $s$-wave
superconductor, the FFLO state is always 1D like and
 no 2D pattern could be obtained, this conclusion is consistent with that of Ref.~\cite{qian,cui}
 and the result will not be presented here.

We start from a phenomenological BCS-type model with the Zeeman
splitting effect caused by an exchange field or in-plane magnetic
field. On a two-dimensional square lattice with a pairing
interaction $V$ between the nearest-neighbor sites, the mean-field
Hamiltonian leading to the $d$ wave superconductivity can be written
as,
\begin{eqnarray}
H&=-\sum_{ ij
\sigma}(t_{ij}c^{\dagger}_{i\sigma}c_{j\sigma}+h.c.)-\sum_{i\sigma}(\mu+\sigma
h)c^{\dagger}_{i\sigma}c_{i\sigma}\nonumber\\& +\sum_{
ij}(\Delta_{ij}c^{\dagger}_{i\uparrow}c^{\dagger}_{j\downarrow}+h.c.),
\end{eqnarray}
where $t_{ij}$ are the hopping constants and $\mu$ is the chemical
potential. $\sigma h$ is the Zeeman energy term, caused by the
interaction between the magnetic field and the spins, with
$\sigma=\pm 1$ representing for spin-up and spin-down electrons,
respectively. The $d$-wave SC order parameter has the following
definition: $\Delta_{ij}=V\langle
c_{i\uparrow}c_{j\downarrow}-.c_{i\downarrow}c_{j\uparrow}\rangle/2$.

This Hamiltonian can be diagonalized by solving the BdG equations,
\begin{equation}
\sum_j \left( \begin{array}{cc}
 H_{i j} & \Delta_{i j}  \\
 \Delta^{*}_{i j} & -H^{*}_{ij}
\end{array}
\right) \left( \begin{array}{c}
u^{n}_{j\uparrow}\\v^{n}_{j\downarrow}
\end{array}
\right) =E_n \left( \begin{array}{c}
u^{n}_{i\uparrow}\\v^{n}_{i{\downarrow}}
\end{array}
\right),
\end{equation}
where $H_{ij}$ is expressed by,
\begin{equation}
H_{ij}=-t_{ij}-(\mu+\sigma h)\delta_{ij}.
\end{equation}

The SC order parameter and the local electron density $n_{i}$
satisfy the following self-consistent conditions,
\begin{eqnarray}
\Delta_{ij}=\frac{V_{ij}}{4}\sum_n
(u^{n}_{i\uparrow}v^{n*}_{j\downarrow}+u^{n}_{j\uparrow}v^{n*}_{i\downarrow})\tanh
(\frac{E_n}{2K_B T}),
\end{eqnarray}
\begin{eqnarray}
n_{i}&=&\sum_n |u^{n}_{i\uparrow}|^{2}f(E_n)+\sum_n
|v^{n}_{i\downarrow}|^{2}[1-f(E_n)].
\end{eqnarray}
Here $f(x)$ is the Fermi distribution function. We define the
ferromagnetic (FM) spin order $m_i$ and the on-site order parameter
$\Delta_i$ as, $m_i=n_{i\uparrow}-n_{i\downarrow}$; $\Delta_i=1/4
(\Delta_{i,i+\hat {x}}+\Delta_{i,i-\hat
{x}}-\Delta_{i,i+\hat{y}}-\Delta_{i,i-\hat{y}})$.

The LDOS is expressed by,
\begin{equation}
\rho_{i}(\omega)=\sum_{n}[|u^{n}_{i\uparrow}|^{2}\delta(E_n-\omega)+
|v^{n}_{i\downarrow}|^{2}\delta(E_n+\omega)],
\end{equation}
where the delta function $\delta(x)$ is taken as
$\Gamma/\pi(x^2+\Gamma^2)$, with $\Gamma=0.01$. The supercell
technical is used to calculate the LDOS.

In the following calculation, we take the hopping constant $t_{ij}$
to be unity for nearest neighbors and zero otherwise. The pairing
potential $V$ and the filling electron density $n$ are chosen as
$V=1.3$ and $n=0.84$ (hole-doped samples with doping $\delta=0.16)$,
respectively. The calculation is made on $48\times 48$ lattice with
periodic boundary condition and random distributed initial values of
the order parameters are chosen. The $10\times 10$ supercell is used
to calculate the LDOS.

\begin{figure}
\centering
  \includegraphics[width=2.6in]{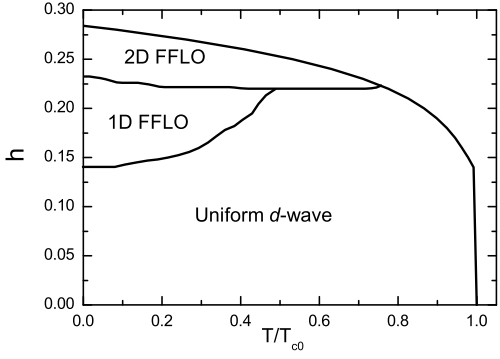}
\caption{H-T phase diagram of the two-dimensional $d$-wave
superconductor in the parallel magnetic field.}
\end{figure}

We summarize our main results in Fig.1, as seen, the $H-T$ phase
diagram is plotted. At zero temperatures two critical Zeeman fields
$h_1=0.14$ and $h_2=0.23$, are revealed. The whole SC state is
divided to be three regions, namely, uniform $d$-wave SC state, 1D
FFLO state, and 2D FFLO state, respectively. The periodicity will
decrease as the magnetic field increases in both 1D and 2D FFLO
states. As $h>0.28$, the SC phase will be destroyed completely. The
periodicity of the 1D FFLO state also increases as the temperature
increases, and it will transit to the uniform state as the
temperature increases further. As a result, the range of the FFLO
phase will decrease upon increasing the temperature. Near the SC
transition temperature, only uniform $d$-wave phase and 2D FFLO
phase was observed, with the transition field at about $h=0.225$.

\begin{figure}
\centering
  \includegraphics[width=2.7in]{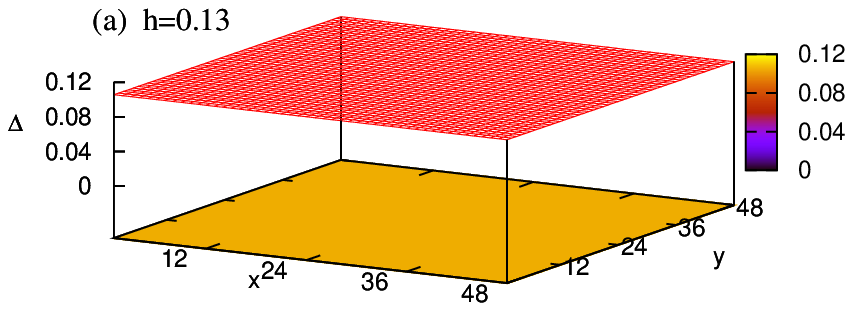}
  \includegraphics[width=2.7in]{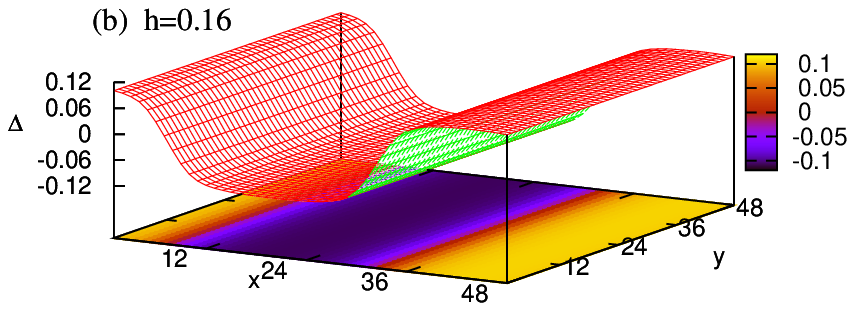}
  \includegraphics[width=2.7in]{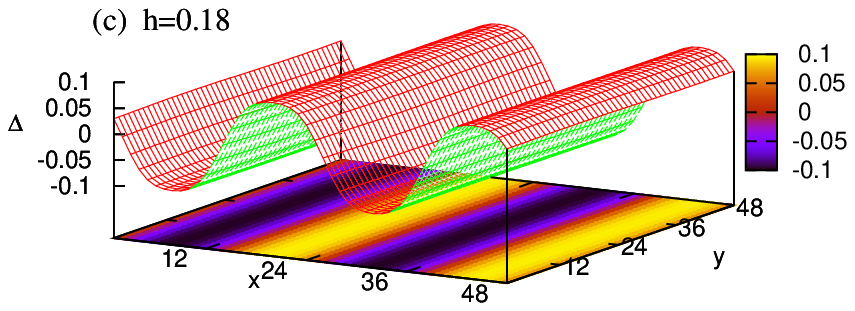}
  \includegraphics[width=2.7in]{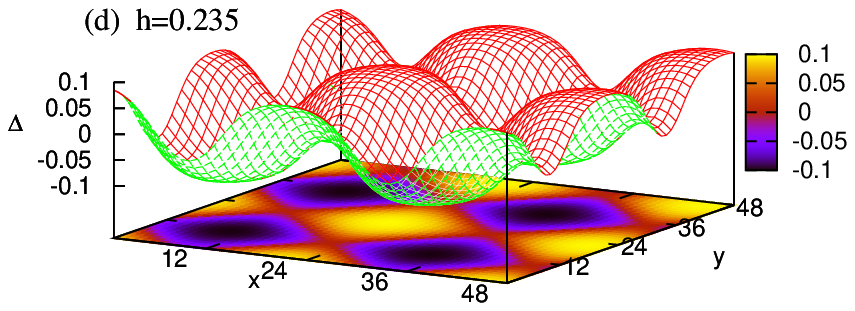}
  \includegraphics[width=2.7in]{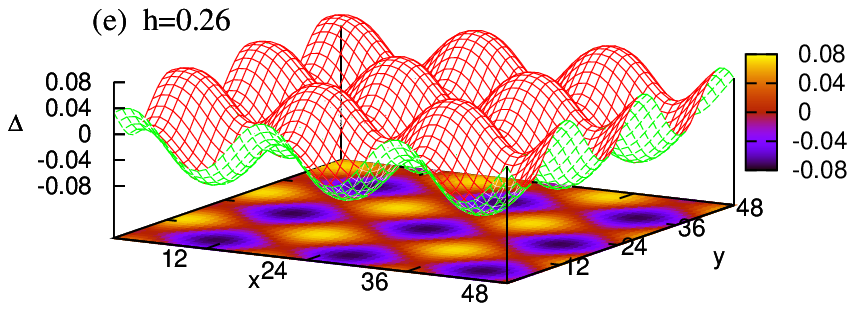}
\caption{(Color online) Plots of the order parameter $\Delta$ as a
function of position for various Zeeman fields $h$.} \label{fig2}
\end{figure}

The calculated order parameter amplitudes for various Zeeman fields
 $h$ with the temperature $T=10^{-5}$ are shown in Figs.2(a)-2(e). As seen, for weaker magnetic field, the
order parameter is uniform [Fig.2(a)]. When the zeeman field
increases, as we can see from Figs.2(b) and 2(c), the SC order forms
the stripe pattern. The order parameter is of nearly cosine form
with the periodicity of about $48$ along $x$ direction as $h=0.16$.
We have verified numerically that the periodicity is kept to be 48
for $0.14<h<0.175$. And the periodicity reduces to 24 as $h$
increases ($0.175<h<0.23)$. Here the finite size effect prevents us
from obtaining solutions with periodicity not commensurate with the
lattice size. As $h$ increases further, the pattern changes to
two-dimensional, with the periodicity decreases as $h$ increases,
which can be seen clearly from Figs.2(d) and 2(e).

\begin{figure}
\centering
  \includegraphics[width=2.7in]{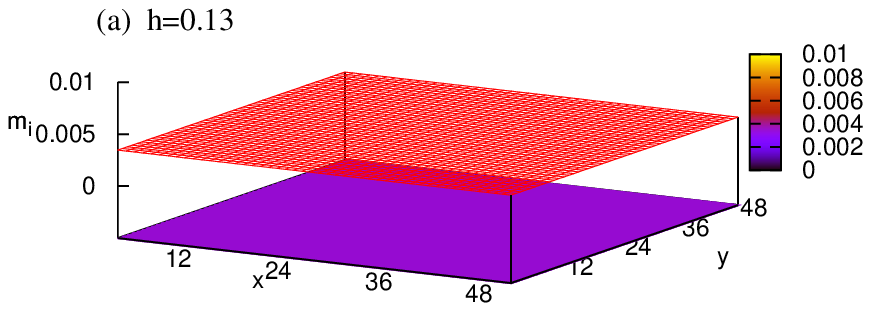}
  \includegraphics[width=2.7in]{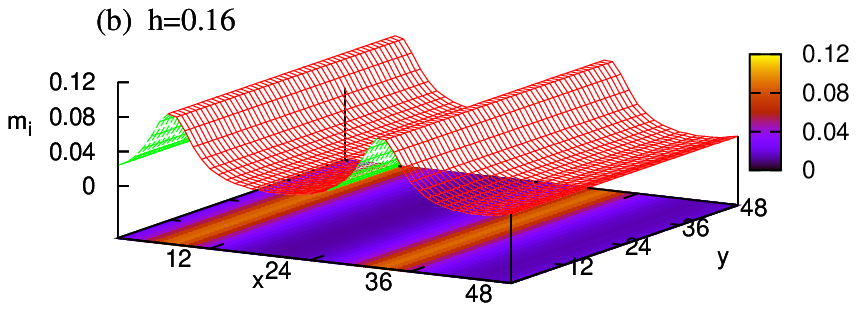}
  \includegraphics[width=2.7in]{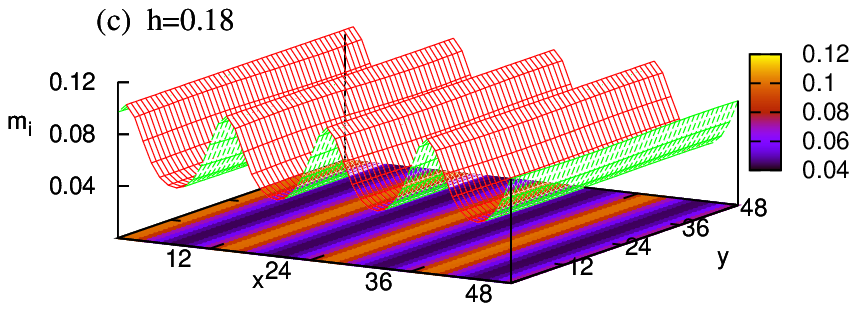}
  \includegraphics[width=2.7in]{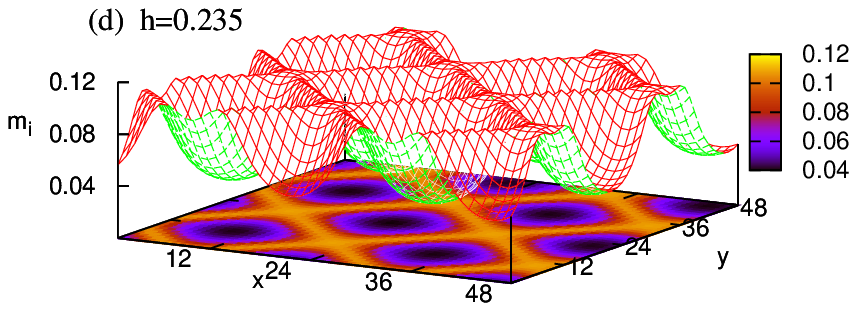}
    \includegraphics[width=2.7in]{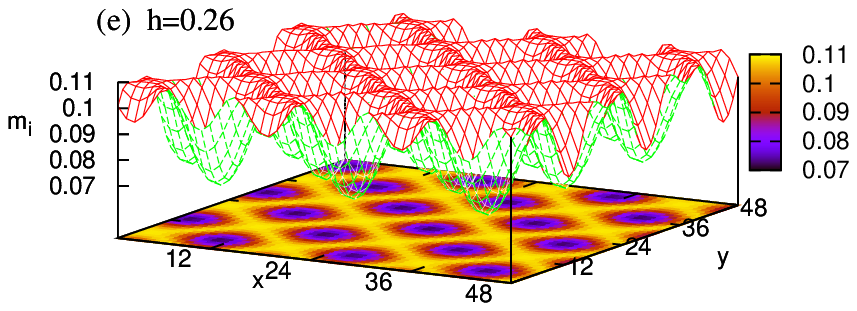}
\caption{(Color online) Plots of the FM order $m_i$ as a function of
position for various Zeeman fields $h$.} \label{fig2}
\end{figure}

The spatial distributions of the FM order are shown in
Figs.3(a)-3(e). The parameters are the same to those in Fig.2. As
seen in Fig.3(a), in the uniform phase, the FM order is also
uniform. Actually, the FM order competes with the SC order and is
suppressed strongly by the SC order, as a result, the FM order could
not survive at lower magnetic field $(h<0.12)$, and quite weak
$(\approx0.003)$ as $h=0.13$. We also checked numerically (not
presented here) that the FM order will increase to about 0.05 in the
normal state for the same magnetic field $(h=0.13)$. In the 1D FFLO
state, as seen in Figs.3(b) and 3(c), the FM order is largest along
the nodal lines and is suppressed when the SC order parameter
increases. The FM order reaches the minimum value as the SC order is
maximum. The pattern also forms 1D stripe but the periodicity is
one-half of that of the order parameter. In the 2D FFLO state
[Figs.3(d) and 3(e)], the FM order forms the checkerboard pattern.
Similar to the case of 1D FFLO state, the FM order is largest at the
nodal lines and minimum as the SC order is maximum. The periodicity
along the parallel direction is the same as that of the order
parameter. While the periodicity along the diagonal direction is
only one-half of that of the order parameter.

We now turn to study the LDOS spectra. The LDOS [Eq. (6)] can be
written as $\rho_i=\rho_{i\uparrow}+\rho_{i\downarrow}$. Here
$\rho_{i\uparrow}$ and $\rho_{i\downarrow}$ are respectively the
spin-up and spin-down parts of the LDOS.
 These two parts are
exactly the same if the Zeeman field is absent. In presence of the
Zeeman field, the spin up LDOS shifts to left and the spin down LDOS
shifts to right. In Figs.4(a)-4(f), we plot the two parts of LDOS
separately to discuss the properties of the LDOS. The whole LDOS
spectra are also plotted so that the results can be compared with
scanning tunneling microscopy (STM) experiments.

The LDOS spectra in the uniform phase are shown in Fig.4(a). As
seen, the spin-up LDOS shifts to the left with the mid-gap point
locating at $\omega=-h$. The SC coherent peaks shift to
$\pm\Delta_0-h$. The spin-down LDOS shifts to the right with the SC
coherent peaks at $\pm \Delta_0 +h$. Outside the gap we can seen the
van Hove peak. As a result, the whole LDOS spectrum contains two
stronger peaks at $\pm(\Delta_0+h)$ and two weaker peaks at
$\pm(\Delta_0-h)$. The gap structure at low energies is "U"-shape.
The density of states at zero energy $\rho(0)$ increases linear with
the external field, indicating that the quasiparticle excitations
due to the magnetic fields.

\begin{figure}
\centering
  \includegraphics[width=3.3in]{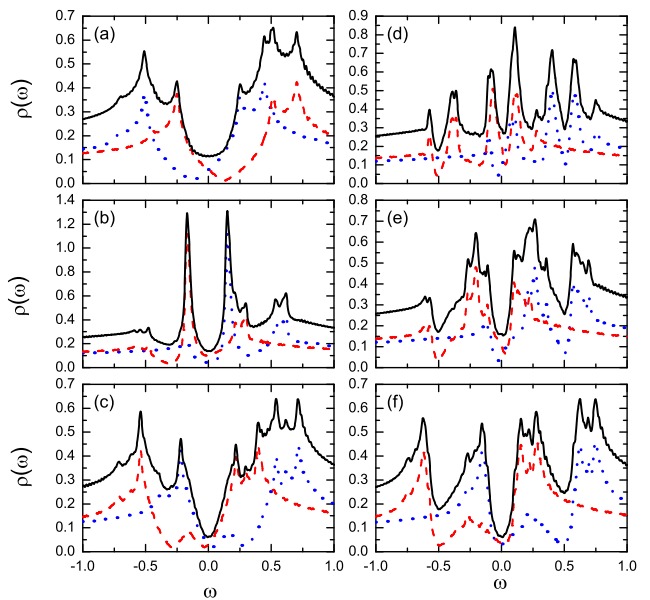}
\caption{(Color online) The LDOS spectra for different phases. Panel
(a) is the LDOS in the uniform phase with $h=0.13$. Panels (b) and
(c) are the spectra in the 1D FFLO phase with $h=0.16$ at the nodal
line and at the site which the order parameter is maximum,
respectively. The right panels are the spectra in the 2D FFLO phase
with $h=0.235$, where (d-f) are the spectra at the saddle point
where two nodal lines intersect, the midsite between two neighboring
saddle points, and the site where the order parameter has the
maximum magnitude, respectively. The (blue) dotted line, (red)
dashed line, and the (black) solid line are spin-up LDOS, spin-down
LDOS, and whole LDOS, respectively.}
\end{figure}

The LDOS spectra in the 1D FFLO phase with $h=0.16$ are shown in
Figs.4(b) and 4(c). Fig.4(b) is for the site on the nodal line. We
can see very sharp and strong peaks at the position $\pm h$. The SC
coherent peaks are suppressed and almost invisible. The peak at
negative energy comes from the spin-up LDOS, and the peak at
positive energy are contributed by the spin-down LDOS. Taking into
account the Zeeman shift, these in-gap peaks (bound states) at $\pm
h$ locate just at the mid-gap position. These bound states are due
to the sign change of the order parameter across the nodal lines and
are related to the Andreev reflections, similar to the mid-gap
states in $d$-wave superconductors~\cite{hu}. The intensity of the
in-gap peaks will decrease as the site moves away from the nodal
line. As we can see from Fig.4(c), at the site where the order
parameter is maximum, the in-gap peaks are turned to be a hump at
the mid-gap position for both spin-up and spin-down LDOS spectra.
The SC coherent peaks are seen clearly. The mid-gap hump is so weak
that it is concealed in the whole LDOS $(\rho_i)$ spectrum. We can
see four peaks at the energies $[\pm (\Delta_0 \pm h)]$. And the
spectrum of the whole LDOS is similar to that of the uniform phase
while the gap structure at low energies is not "U"-shape but
"V"-shape due to the presence of the mid-gap hump.

At last we plot the LDOS spectra of the 2D FFLO phase in
Figs.4(d)-4(f). Actually the features of the spin-up LDOS spectra
are studied intensively in Ref.~\cite{qian}. There are two kinds of
Andreev bound states. One is due to the sign change of the order
parameter across the nodal lines. The second is essentially
localized at the saddle points. The order parameter is suppressed
strongly in a intersecting region, which produces a potential well
for a quasiparticle and thus generates two finite-energy andreev
bound states. As a result, at the saddle points, four in-gap peaks
exist in the spin-up LDOS spectra [Fig.4(d)] at the energies
$-0.575$, $-0.385$, $-0.08$ and $0.105$. And mid-gap peaks exist
between two neighboring saddle points [Fig.4(e)]. At the site where
the order parameter is maximum, the LDOS spectrum [Fig.4(f)] is
similar to that of the 1D FFLO state [Fig. 4(b)] and that of the
uniform phase, namely, if the van hove peaks and the weak peak
caused by the mid-gap hump are excluded, there are only four peaks
left, locating at $\pm(\Delta_0\pm h)$, contributed by the spin-up
and spin-down LDOS, respectively.

We have shown the LDOS spectra of the three different phases. As
seen in Figs.4(a)-4(f), the spectra are quite different and the
spectra in each phase have their distinctive features as we
discussed above. Thus they can be easily detected by the STM
experiments and can be used as signatures to probe the FFLO states.

In summary, based on a BCS-type model and BdG equations, we studied
the phase transition induced by the external magnetic field. The
phase diagram is mapped out and the transitions from uniform phase
to 1D FFLO state, and 1D FFLO state to 2D FFLO state are revealed.
We also calculate the LDOS to discuss the signatures of the three
phases, namely, the LDOS spectra in the uniform phase will contain
four peaks due to the Zeeman shift. In the 1D FFLO state, the LDOS
spectra show mid-gap states due to the Andreev reflection. In the 2D
FFLO states, four in-gap peaks are revealed at the saddle point due
to two kinds of Andreev bound states.

This work was supported by the Texas Center for Superconductivity at
the University of Houston and by the Robert A.Welch Foundation under
the Grant no. E-1146.

\end{document}